\documentclass{emulateapj}

\slugcomment{To appear in ApJ}

\shorttitle{Panchromatic view of \objectname{M94 (NGC4736)}}
\shortauthors{Trujillo et al.}

\begin{document}


\title{Unveiling the nature 
of \objectname{M94's (NGC4736)} outer region: a panchromatic perspective}


\author{Ignacio Trujillo\altaffilmark{1}, Inma Martinez-Valpuesta and David Mart\'inez-Delgado\altaffilmark{1}}
\affil{Instituto de Astrof\'isica de Canarias, E-38205, La Laguna, Tenerife, Spain}
\affil{Departamento de Astrofísica, Universidad de La Laguna, E-38205 La Laguna, Tenerife, Spain}
\email{trujillo@iac.es}
\author{Jorge Pe\~narrubia}
\affil{Institute of Astronomy, University of Cambridge, Madingley Road, Cambridge CB3 0HA, UK}
\author{R. Jay Gabany}
\affil{Blackbird Observatory, New Mexico, USA}
\and
\author{Michael Pohlen}
\affil{Cardiff University, School of Physics \& Astronomy, Cardiff, CF24 3AA, Wales, UK}


\altaffiltext{1}{Ram\'on y Cajal Fellow}


\begin{abstract}

We have conducted a deep multi-wavelength analysis (0.15--160 $\mu$m) to study the outer region
of the nearby galaxy M94. We show that the non-optical data support the idea that the outskirts
of this galaxy is not formed by a closed stellar ring (as traditionally claimed in the
literature) but by a spiral arm structure. In this sense, M94 is a good example of a Type III
(anti-truncated) disk galaxy having a very bright outer disk. The outer disk of this galaxy
contains $\sim$23\% of the total stellar mass budget of the galaxy and contributes  $\sim$10\%
of the new stars created showing that this region of the galaxy is active. In fact, the specific
star formation rate of the outer disk ($\sim$0.012 Gyr$^{-1}$) is a factor of $\sim$2 larger
(i.e. the star formation is more efficient per unit stellar mass) than in the inner disk. We
have explored different scenarios to explain the enhanced star formation in the outer disk. We
find that the inner disk (if considered as an oval distortion) can dynamically create a spiral
arm structure in the outer disk which triggers the observed relatively high star formation rate
as well as an inner ring similar to what is found in this galaxy.

\end{abstract}





\section{Introduction}

Recent years have seen a revolution in our understanding of the outskirts of spiral galaxies. On the one
hand, UV emission in the disk of spiral galaxies has been discovered at very large radii (the
so-called XUV disks Thilker et al. 2005; Gil de Paz et al. 2005; see also Donas et al. 1981),
well beyond the optical radius.  On the other hand,  new studies have  challenged the
traditional picture of disks of spirals following an exponential radial surface brightness
profile (Freeman 1970) down to a large radius where a truncation happens (van der Kruit
1979; van der Kruit \& Searle 1981a,b). In fact, recent systematic observations show an unexpected
diversity of behaviors in the faintest regions of the surface brightness profiles.   Three
main categories or Types are now established: Type I, in which the disk does in fact show a
simple exponential form; Type II where the inner part of the disk shows an exponential
fall--off followed by a steeper profile, and Type III, where the inner exponential profile is
followed by a shallower, often exponential, profile (Pohlen et al. 2002; Bland-Hawthorn et al. 2005;  Erwin et
al. 2005, 2008; Hunter \& Elmegreen 2006; Pohlen \& Trujillo 2006). There is evidence (Pohlen
\& Trujillo 2006; Erwin et al. 2007) that these three types are related with the global morphology
of the galaxies (i.e. Type II are more common in later types), implying a suggestive link
between the evolutionary patterns followed by the galaxies and the distribution of stars in
their outskirts.

Previous observational results suggest that the outer disks provide a direct view of disk
assembly. For this reason, several theories have tried to explain the above variety of
behaviors in the outer regions of galaxies. Among them we can mention ideas from Yoshii \&
Sommer-Larsen (1989) who found that to explain Type I profiles the time scales of both
viscosity and star formation through the disks should be comparable (see also, Ferguson \&
Clarke 2001 and Slyz et al. 2002). Also it is possible to find ideas based on star formation
thresholds (Kennicutt 1989; Elmegreen \& Parravano 1994; Schaye 2004; Elmegreen \& Hunter
2006), the maximum angular momentum  of the gas that is collapsing to form the disk (van der
Kruit 1987; van den Bosch 2001), or secular evolution  driven by bars/spiral arms or clumps
disruption (Debattista et al. 2006;  Bournaud et al. 2007; Ro\v skar et al. 2008, Foyle et
al. 2008) to explain the Type II morphology. To justify Type III, tidal stripping caused by
minor galaxy mergers (Pe\~narrubia et al. 2006; Younger et al. 2007), dark matter subhalos
bombarding (Kazantzidis et al. 2008) or high eccentricity flyby by a satellite galaxy
(Younger et al. 2008)  have been proposed. Although these theories qualitative explain the
observed behaviors, a quantitative, detailed analysis of the observations that allow us to
judge between the different scenarios is missing. 

In this paper, we will focus on the nearby galaxy \objectname{\objectname{M94 (NGC4736)}}. As
we will show through the paper, this galaxy is an example of the brightest anti-truncated (Type
III) outer disks that can be found in nearby galaxies.  The vicinity of the object (4.66 Mpc;
Karachentsev 2005) makes it a perfect candidate to explore in great detail the properties of
its outer disk, and consequently, shed some light on the nature of the extended disk
phenomenon, particularly, for the most brightest (anti-truncated) outer disk cases.  

To achieve this goal we have compiled data from UV (GALEX), optical (SDSS), near-infrared
(2MASS) and infrared (Spitzer) to obtain the surface brightness distributions down to faint
surface brightness in 17 different filters. The remainder of the paper is organized as follows.
In Section 2, we give a brief summary of the structural characteristics of M94. Section 3
contains a detailed description of the data that we have used to conduct our analysis. The
dependence with the wavelength of the outer galaxy structure is shown in Section 4. Section 5
shows the surface brightness profiles of the galaxies in 17 different filters from FUV to MIPS
160$\mu$m. The results of our analysis are explained in Section 6. Different formation scenarios
are proposed to explain our results in Section 7. Finally, in Section 8 we summarize our main
results.

\section{M94 (NGC4736) in a nutshell}

\begin{figure}
\includegraphics[angle=-90,scale=0.65]{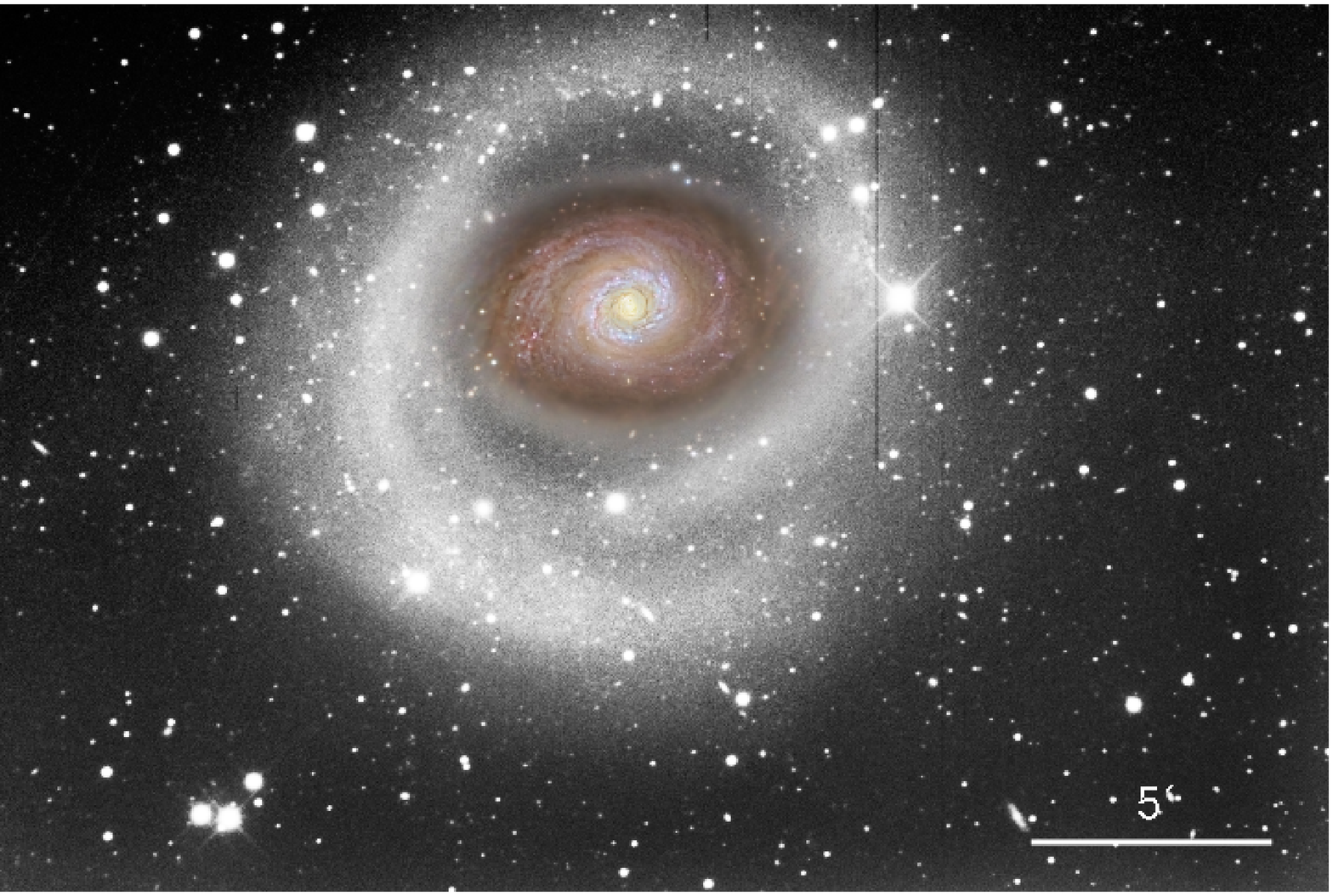}

\caption{Luminance clear-filter very deep image of M94 (NGC4736) obtained with the BBRO
0.5-meter telescope. The total exposure time of the image was 555 min. The image size is
27.7\arcmin$\times$18.2\arcmin. North is up and East is left. For illustrative purpose, a
color image of the central part of the galaxy obtained with the same telescope has been
superimposed on the inner (saturated) disk region of the galaxy. It is worth stressing that
the traditional ("textbook") picture of M94 is just what it is shown in color here.}

\label{luminance}

\end{figure}

M94 is the closest early-type spiral galaxy  with low inclination (see Fig. \ref{luminance}).
It is located in the Canes Venatici I cloud and  was classified as (R)SA(r)ab-type according to
de Vaucouleurs et al. (1991) and reclassified as (R)SAB(rs)ab-type by Buta et al. (2007). Its
total apparent magnitude in the B-band is 8.54 mag, its R$_{25}$ is $\sim$337\arcsec  \ and it
is claimed to be at an inclination of 35 degrees (Erwin 2004).

Up to five main regions have   been identified in this system according to Bosma et al. (1977):
a bulge R$<$15$\arcsec$  (see also, Beckman et al. 1991),  an inner ring of ongoing starburst
activity (R$\approx$45$\arcsec$; van der Kruit 1976), an oval (disk/bar) stellar distribution
down to intermediate distance (R$\approx$220$\arcsec$; Beckman et al. 1991; Shaw et al. 1993;
M\"ollenhoff et al. 1995; Mulder 1995),  a zone of low surface brightness (a gap), and what
has been traditionally considered as an outer stellar ring around R$\approx$350$\arcsec$ (see
also, Buta 1988; Mu\~noz-Tu\~n\'on et al. 1989). Fig. \ref{luminance} illustrates the different
regions of this galaxy as viewed by a very deep exposure obtained by our group (see
details in the next section).

M94 is also known  for hosting the closest example of a low ionization nuclear emission line
(LINER) 2 nucleus (see e.g. Roberts et al. 2001), although the origin of this emission as
powered by a low-luminosity active galactic nucleus (LLAGN) is uncertain. Several observations
supporting the hypothesis of a non-stellar  origin for this emission are the detection of a strong
non-thermal radio continuum source at the position of the nucleus (Turner \& Ho 1994), the
detection in UV using the HST of two bright point sources  in the nuclear region with possible
bow shocks resulting from an hypothetical merger of two supermassive blackholes (Maoz et al.
1995), and the detection of a X-ray spectrum dominated by a Seyfert-like power-law continuum
(Roberts, Warwick \& Ohashi 1999). On the other hand, favoring the stellar emission hypothesis
to explain the LINER emission are the large [FeII]/Pa$\beta$ ratio (Larkin et al. 1998),  and
the H$\alpha$ (Taniguchi et al. 1996) and FIR luminosities (Smith et al. 1991; 1994).

\section{Observations}

The analysis conducted in this paper is based on a multi-wavelength coverage of the galaxy.
Consequently, the data comes from a variety of facilities and has been mostly obtained from
archives. Below follows a brief
description of the data collected.

\subsection{Deep optical luminance imaging}

In order to explore in detail the structural characteristics of the outer region of
M94 in the optical regime, we decided to take a very deep exposure of
this galaxy (see Fig. \ref{luminance})  using the Ritchey-Chr\'etien 0.5-meter telescope of the
BlackBird Remote Observatory (BBRO) situated in the Sacramento Mountains (New Mexico, USA).
The camera used was a Santa Barbara Instrument Group (SBIG) STL-11100 CCD camera, which has a
FOV of 27.7\arcmin$\times$18.2\arcmin and a pixel size of 0.45\arcsec. The angular
resolution of this image is $\sim$3.4\arcsec. 

To maximize the photon collection our imaging consists on multiple deep exposures through a
very broad non-infrared clear luminance filter (3500$<$$\lambda$$<$8500 \AA), and a set of
red, green and blue filters from the SGBI custom scientific filters
set\footnote{http://www.sbig.com/products/}. The total luminance exposure time allocated to
this image was 555 min. In addition, to decrease the effect of noise, and consequently
increase the detection and enhancing of the faint structures we have applied a Gaussian blur
filter (Davies 1990; Haralick \& Shapiro 1992). The sigma of the gaussian kernel used on
convolving the images is 1 pixel. This kernel was applied only to the outer regions to reveal
more clearly this structure. As shown in a previous work (Mart\'inez-Delgado et al. 2008) the
faintest structures reachable with the BBRO telescope using the luminance filter with 
similar exposure times is $\gtrsim$27.5 mag/arcsec$^2$ in the R-band. 

Since we lack a calibration of the optical luminance filter, our image is used only to show
the faint structure of the outer disk at these wavelengths. 

\subsection{GALEX UV}

The ultraviolet observations are obtained from the GALEX (Galaxy Evolution Explorer) mission
(Martin et al. 2005). In particular, these data were retrieved from the Nearby Galaxy Survey
mission (Bianchi et al. 2003a,b; Gil de Paz et al. 2004; 2007). GALEX imaging benefits from a
large field-of-view  (1.25$^\circ$ diameter), a very low sky background and high sensitivity.
We have used its two ultraviolet filters: a far-ultraviolet (FUV) filter centered
at 1528 \AA (FWHM 269 \AA) and a near-ultraviolet (NUV) filter centered at 2271 \AA (FWHM 616
\AA). The angular resolution of these imaging is $\sim$4.6\arcsec, and pixel size is
1.5\arcsec. Absolute calibration uncertainties are $\sim$15\% in both the far- and
near-ultraviolet (Dale et al. 2007). An image of \objectname{M94 (NGC4736)} in the ultraviolet filters can be seen
in the left panel of Fig. \ref{colmosaic}.

\subsection{SDSS  optical}

Optical data in 5 bands (u',g',r',i' and z') were obtained from the Sloan Digital Sky Survey
(SDSS; York et al. 2000) archive. SDSS images have a pixel size of 0.396\arcsec  and typical
angular resolution of $\sim$1\arcsec. Due to the large apparent size of the galaxy we had to
create a mosaic of 8
individual SDSS frames (13.5\arcmin$\times$9.8\arcmin each) to cover the total
galaxy angular extension. An image of the galaxy in the SDSS optical bands can be seen in the
central panel of  Fig. \ref{colmosaic}.  

\subsection{2MASS near-infrared}

J$_s$, H and K$_s$ images were extracted from the 2MASS  NASA/IPAC Infrared Science Archive
(IRSA). This galaxy is included in the 2MASS Large Galaxy Atlas (Jarrett et al. 2003). This
atlas already provides a single image (20\arcmin$\times$23.3\arcmin) covering this galaxy
composed of the individual 2MASS fields. The angular resolution is $\sim$3\arcsec and the pixel
size is  1\arcsec.

\subsection{Spitzer infrared}

IRAC (3.6$\mu$m, 4.5$\mu$m, 5.8$\mu$m and 8$\mu$m) bands and MIPS (24$\mu$m, 70$\mu$m, and
160$\mu$m) bands were obtained from the Spitzer Legacy program SINGS (Kennicutt et al. 2003).
The pixel size is 0.75\arcsec for the IRAC (24.6\arcmin$\times$17.9\arcmin) mosaics, and
1.5\arcsec, 4.5\arcsec and 9\arcsec for MIPS 24$\mu$m, 70$\mu$m and 160$\mu$m mosaics
respectively.  Calibration uncertainties are $\sim$10\% for IRAC data, and 4\% (24$\mu$m), 7\%
(70$\mu$m), 12\% (160$\mu$m) for MIPS data (Dale et al. 2007). An image of \objectname{M94
(NGC4736)} in the Spitzer infrared filters can be seen in the right panel of Fig.
\ref{colmosaic}. The angular resolution of these images is summarized in Table 1.

\section{Outer galaxy structure wavelength dependence}

\begin{figure}
\includegraphics[angle=0,scale=0.30]{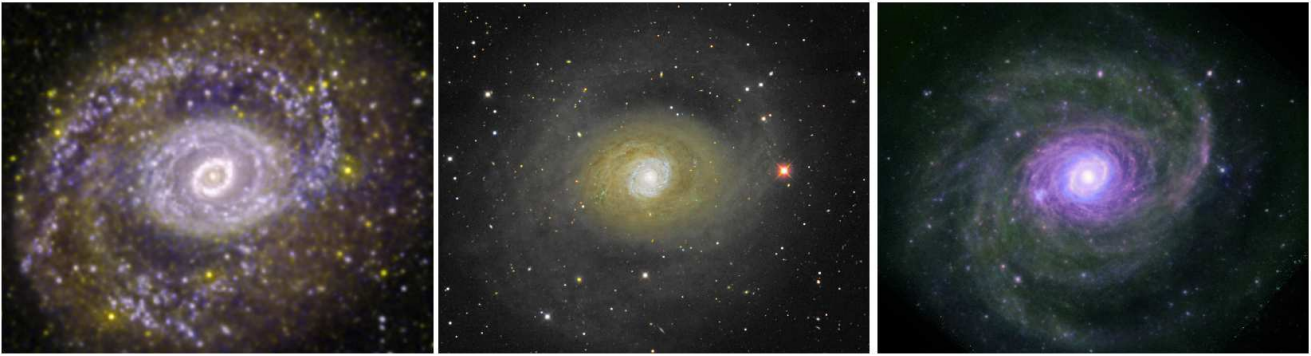}

\caption{Panel showing M94 (NGC4736) galaxy through the ultraviolet, optical and infrared
wavelengths. Left panel shows M94 as viewed by the GALEX bands (blue:
FUV; yellow: NUV). Middle panel shows M94 as viewed by the SDSS bands
(blue: g-band; green: r-band; red: i-band). Right panel shows M94 as
viewed by the Spitzer bands (blue: 3.6$\mu$m; green: 8$\mu$m; red: 24$\mu$m). In all the
cases North is up and East is left. The area shown (18\arcmin$\times$15\arcmin) is selected
for being the  common area covered by the set of facilities used in this work.}

\label{colmosaic}

\end{figure}

The outer region of \objectname{M94 (NGC4736)}  is claimed to be a  stellar ring, as mentioned
in Section 2. This is in fact the appearance of this structure when observed at the optical
wavelengths (see the optical luminance image in Fig. \ref{luminance} and the optical SDSS image
in the central panel in Fig. \ref{colmosaic}). The nature of this region, however, seems to be
radically different when it is probed both in the UV and in the infrared (see left and right
panels in Fig. \ref{colmosaic}). In fact, at these wavelengths, the outer region is better
described as an outer disk having spiral arms.

Further support for the outer region being a disk with a spiral arm structure and not a closed
stellar ring can be found from the deep  HI imaging done by Braun (1995; his Fig. 8) and de Blok
et al. (2008; their Fig. 80). In that
Figures, a bright spiral arm is seen in the outer disk. This arm, that is not visible in the
optical, corresponds to the brightest feature observed in the outer region on the Spitzer
24$\mu$m image (see Fig. \ref{colmosaic}) and it is also very clear in the UV imaging.
Consequently, the non-optical data strongly points out in the direction of that the outer region
of M94 is actually a disk with a spiral arm structure and not a ring as has been traditionally
claimed in the literature. On what follows we explore the properties of this outer disk in
comparison to the different regions of the galaxy.

\section{Surface brightness profiles}

In order to derive surface brightness profiles in a consistent way we use the same set of
ellipticity and position angle in all our bands. Following Pohlen \& Trujillo (2006) we explore
which value of the ellipticity and position angle (PA) are a good representation of the outer disk.
To remove as much as possible potential effects related to star formation and dust extinction we use
the 3.6$\mu$m image as a proxy. Briefly, the method works as follows, we use the IRAF task
\textit{ellipse} (STSDAS package) on the sky substracted and masked (from all non-galaxy component)
image.  The center of the ellipses is fixed by a gaussian fit to the bright nucleus of the galaxy.
We used a logarithmic radial sampling with steps of 0.03 to increase the S/N especially in the outer
parts. Iterative 3$\sigma$ rejection along the ellipse is applied to minimise the influence of
cosmic rays or remaining foreground stars. The free ellipse fit (fixed center, free ellipticity and
position angle) is used to determine the best set of ellipticity and PA describing the outer disk.
These values are taken at the radius where the mean flux of the best fitted free ellipse reaches the
value of the standard deviation of our background sky. This limit ensures enough S/N to fit a free
ellipse but is small enough to be in the radial region dominated by the outer disk.

Using 3.6$\mu$m image, we find that the outer isophotes are well described by an ellipticity
of 0.14 and a PA of 122.2 degrees (anticlockwise from North). This ellipticity implies
an inclination of $\sim$8 degrees for the galaxy, far away from previous reported values
of 35 degrees based on the shape of the inner (oval distortion) disk. The PA is also
different from the one reported previously of 108 degrees (Sofue et al. 1999) for the
same reason.

The surface brightness profiles in all the filters were obtained using a task
custom-made  using IDL. Bright stars were identified and masked. We measured
the median intensity and rms along  elliptical apertures with fixed ellipticity and
position angle. Our code was tested against the IRAF task \textit{ellipse} producing
very similar results. 

\begin{figure}

\includegraphics[angle=0,scale=1]{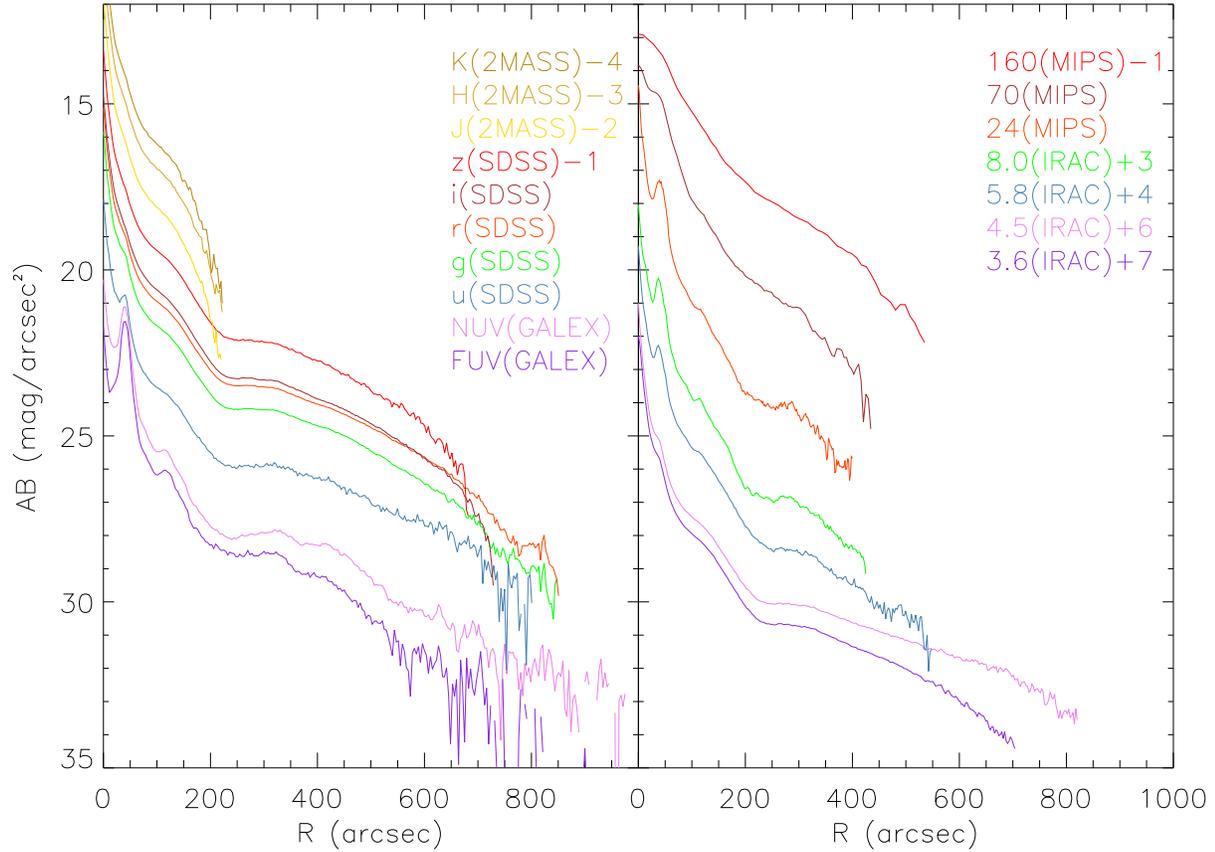}

\caption{Surface brightness profiles (AB system) of M94 (NGC4736) in 17 bands ranging from
the FUV to MIPS 160$\mu$m.}

\label{profiles}

\end{figure}

A critical issue for obtaining accurate surface brightness profiles at the low surface brightness
levels of the galaxies is the accuracy of the sky substraction. Although the archive data used in
this work is sky subtracted we repeated the "ellipse method" explained in Pohlen \& Trujillo (2006)
to improve the sky determination. In brief, the idea is to expand the surface brightness estimation
using the ellipticity and PA obtained from the outer disk. By plotting the flux at these ellipses as
a function of radius it becomes clear at which radial distance the ellipses leave the galaxy and
enter the background, flat flux (noise), region (see Fig. 2 of Pohlen \& Trujillo 2006). The mean
value and standard deviation of the fluxes within the above radius and the radius extended by 20\%
are used as the final sky values. This sky value is subtracted to the original image and the process
of estimating the surface brightness is repeated. Our surface brightness estimation (given in the AB
system) are shown on Fig. \ref{profiles}.

Each image has a critical surface brightness ($\mu_{crit}$) beyond that the reported surface
brightness profiles are not reliable. Using SDSS imaging, Pohlen \& Trujillo (2006) defined this
$\mu_{crit}$ as the value where the profiles obtained by either over- or undersubtracting the sky by
$\pm$1$\sigma$-sky start to deviate by more than 0.2 mag. This value is  0.54 mag above the limiting
surface brightness ($\mu_{lim}$) of the image calculated as the 3$\sigma$ error of the sky value
determination. Table \ref{surcrit} compiles the set of $\mu_{crit}$ for the different bands used in
this work. The surface brightness are corrected for Galactic extinction (Schlegel et al. 1998) using
E(B-V)=0.018 mag. No attempt was made to correct the surface brightness measurements for internal
extinction, since no unique recipe is available to do this. Consequently, this will introduce an
additional uncertainty on the surface brightness estimation. In addition, the galaxy studied here is
a fairly face-on system (the measured ellipticity implies an 8 degree inclination) so the expected
correction is only very small. In fact, if we adopt a popular parameterization of the effects of
internal extinction (see e.g. Cho \& Park 2009): A$_\lambda$=$\gamma_\lambda$$\log_{10}$(a/b) where
$\gamma_\lambda$ is a constant that depends on the observed wavelength but which value is $\sim$1,
and a/b is the axis ratio, then in our case, $\log_{10}$(1.16)=0.064. So the expected corrections
are really minor.

Following the estimations of $\mu_{crit}$ we find that, with the exception of the shallow 2MASS
near-infrared images (which do not allow us to study the outer disk), we can choose the
following outest radius (430\arcsec) as a compromise among the different profiles depth. From
now onwards, consequently, we will divide the galaxy into the following zones: bulge region
(0-75\arcsec), inner disk region (75-200\arcsec), and outer disk region (200-430\arcsec).  

\section{Results}

\subsection{Surface brightness profile shapes}

The observed profiles coupled with the galaxy morphology suggests classifying this galaxy as a Type
III (Erwin et al. 2005). In fact, one of the most interesting features of the surface brightness
profiles of M94 is the unchanging nature of the global structure of the galaxy from the UV to the
far infrared. This implies that stars of all ages as well as dust are similarly distributed
throughout the galaxy structure. The similarity between the UV and FIR profiles is more or less
expected as both wavelength ranges are tracing the most active star formation regions of the
galaxies. In relation to the optical and NIR data (although we warn that for these last data we lack
the information in the outer region) which trace the intermediate-old stellar component, we think
that the fact that they also share the shape of the young stellar component implies that the galaxy,
globally, is  actively forming new stars through all the disk and that this activity has been
maintained similarly in the last Gyrs. As we will show later in the text, this scenario is supported
by the specific star formation rate distribution of the galaxy which is relatively constant within a
factor of 3 through the whole galaxy and by the fact that the gas exponential infall time scales are
similar along the galaxy.

An exception is seen for the 70 and 160 $\mu$m MIPS bands, where we see the inner star forming
ring is almost invisible. However, that is clearly the effect of the larger PSF at these
wavelengths (cf. Table 1).

One of the most interesting feature of the surface brightness profile of M94 is the brightness
($\sim$23.5 mag/arcsec$^2$ in r'-band) at which the outer disk appears. Normally, for Type III
galaxies, the break occurs at faintest surface brightness. For example, Pohlen \& Trujillo (2006)
find a typical surface brightness of $\sim$24.7$\pm$0.8 mag/arcsec$^2$ in the r'-band, whereas Erwin
et al. (2008) find  $\sim$24.2$\pm$0.5 mag/arcsec$^2$ in the R-band. This makes M94, in comparison,
a Type III galaxy having an outer disk on the very bright side. However, M94 is not an unique
example within the family of Type III objects, in fact, it is possible to find other galaxies with
similar characteristics. For example, M77 (NGC1068) is strikingly similar to M94 (see its profile on
Pohlen \& Trujillo 2006). In the case of M77, the outer disk starts at $\sim$23 mag/arcsec$^2$, and
morphologically also, both have similar features. For this reason, we think that both of these two
galaxies could be considered as examples of  Type III galaxies having the brightest ($<$23.5
mag/arcsec$^2$ in r'-band) anti-truncated outer disks. However, a note of caution should be stated
about the Type of the galaxy M94. The Type III classification is based on the assumption (following
the traditional picture of this galaxy) that the inner region (75-200\arcsec) is a true disk. If
this region is, however, considered as an oval distortion (i.e. a structure with a bar--like nature
and not a real disk), then the genuine disk of the galaxy is the outer region ($>$200\arcsec) and
the galaxy should be better considered as a Type I.

\subsection{Stellar mass surface density profile}

\begin{figure}

\includegraphics[angle=0,scale=1]{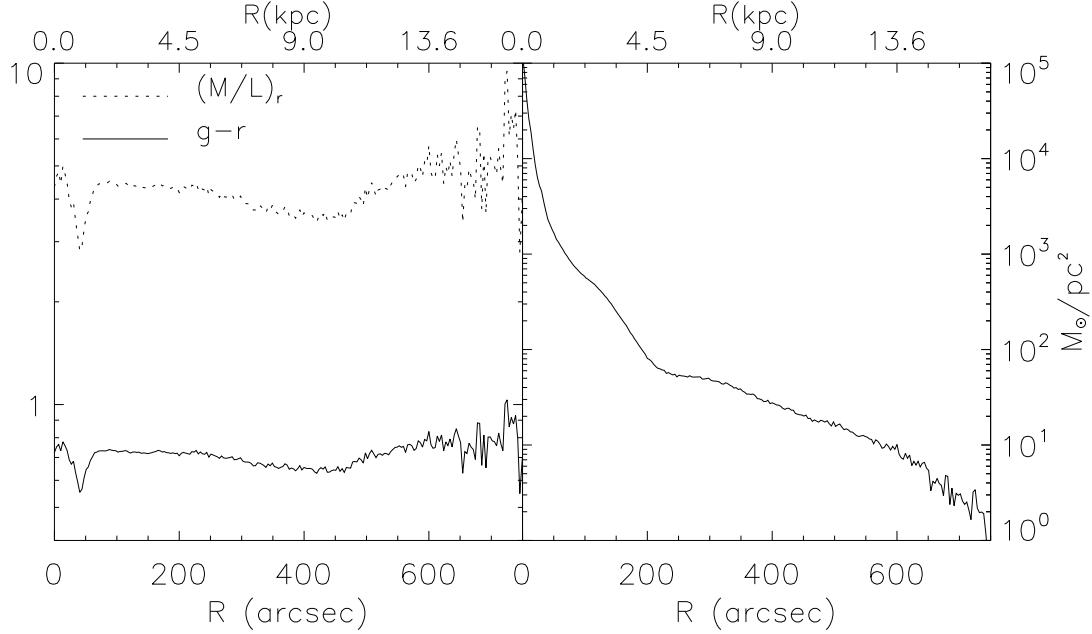}

\caption{Left panel: (g-r) color and (M/L)$_r$ profiles of M94. Right Panel. Stellar mass density
profile of M94 (NGC4736) obtained using the r-band profile and (M/L)$_r$ from the (g-r)
color following the prescription given by Bell et al. (2003). }

\label{massprof}

\end{figure}

At any given wavelength it can be shown that the surface brightness profiles
$\mu_{\lambda}$ can be converted into stellar mass surface density profiles as follows:

\begin{equation}
\log(M/(M_{\sun}pc^2))=\log(M/L)_{\lambda}-0.4(\mu_\lambda-m_{abs,\sun,\lambda})+8.629
\label{massdensity}
\end{equation}

with $(M/L)_{\lambda}$ the stellar mass--to--light ratio in solar units and
$m_{abs,\sun,\lambda}$ the absolute magnitude of the Sun at a given wavelength $\lambda$.
Lacking a deep near-infrared image, we have chosen the SDSS r-band  to estimate the M/L ratio.
This band is selected because it is the reddest of the deepest SDSS bands. To estimate the
(M/L)$_r$ ratio we have used the (g-r) color following the prescription given by Bell et al.
(2003). We add 0.15 dex to the M/L obtained in Bell et al. (2003) to normalize the predicted M/L
to a Salpeter (1955) IMF. The (g-r) color profile, the (M/L)$_r$ and the stellar mass density
profile are shown in Fig. \ref{massprof}. The minimum observed both in color and (M/L)$_r$ is
associated with the nuclear stellar ring. The shape of the stellar mass density profile is very
similar to the shape of the reddest (stellar origin) surface brightness profiles we have. The
stellar mass density at which the outer disk starts is $\sim$60 M$_{\sun}$/pc$^2$. As expected
due to M94's bright outer disk,  this value is higher than the mean value measured for
late--type Type III galaxies ($\sim$10 M$_{\sun}$/pc$^2$; Bakos et al. 2008).

Using the stellar mass density profile based on colors, one can calculate a rough estimate of
the total stellar mass of the galaxy: $\sim$6.5$\times$10$^{10}$ $M_{\sun}$. In each of the
galaxy regions\footnote{We would like to emphasize that the term regions used here (as well as
in the rest of the paper) should not be confused with the actual components of the galaxy (i.e.
bulge, inner disk, outer disk, etc). This distinction is relevant when estimating quantities
like the amount of mass since the real components of the galaxy can indeed overlap.} the amount
of stellar mass we measure is: bulge region (3.2$\times$10$^{10}$ $M_{\sun}$), inner disk region
(1.7$\times$10$^{10}$ $M_{\sun}$) and outer disk region (1.5$\times$10$^{10}$ $M_{\sun}$). Due
to the significant brightness of the outer disk, its contribution to the total stellar mass
budget of the galaxy is quite significant ($\sim$23\%). This is a factor of $\sim$2 larger than
the average late-type Type III outer disk (Bakos et al. 2008).

\subsection{Star formation rates}

To assess the origin of the outer disk structure a key element is to measure its star
formation rate. This value can give us clues on whether this region of the galaxy is
still forming (i.e. creating new stars) or whether it is an old structure. Following
Mu\~noz-Mateos et al. (2007), the star formation rate (SFR) can be computed from the
apparent magnitude (corrected for internal extinction) in the FUV (AB system) using the
calibration given by Kennicutt (1998):

\begin{equation}
\log(SFR)(M_{\sun}yr^{-1})=2\log d(pc)-0.4FUV-9.216
\label{sfrformula}
\end{equation} 

This can be transformed to a  SFR per surface area as follows:

\begin{equation}
\log(SFR_\mu)(M_{\sun}yr^{-1}pc^{-2})=-0.4\mu_{FUV}+1.413
\label{musfr}
\end{equation} 

To estimate the internal extinction A$_{FUV,in}$ in the FUV, we follow the empirical
relation between the TIR-to-FUV ratio (i.e. the total luminosity in the infrared over
the luminosity in the FUV), $\log(F_{TIR}/F_{FUV})$ and the observed (FUV-NUV) color
derived by Boissier et al. (2007). Once $\log(F_{TIR}/F_{FUV})$ is obtained the
attenuation profiles are computed using the fit of Buat et al. (2005):

\begin{equation}
A_{FUV,in}=-0.0333X^3+0.3522X^2+1.1960X+0.4967
\end{equation}

with X=$\log(F_{TIR}/F_{FUV})$. The median value of A$_{FUV,in}$ within the inner 430$\arcsec$ is
1.7 mag. This translates (using  A$_K$=0.0465A$_{FUV}$; Mu\~noz-Mateos et al. 2007) to an inner median
extinction in the K-band profile of 0.079 mag.

Using Eq. \ref{sfrformula} we can estimate the total SFR of the galaxies as well as the SFR at
the different components. We find that the total SFR for M94 is 1.04 $M_{\sun}yr^{-1}$ which is
distributed across the different regions of the galaxy as follows: bulge (0.75
$M_{\sun}yr^{-1}$), inner disk (0.14 $M_{\sun}yr^{-1}$) and outer disk (0.15 $M_{\sun}yr^{-1}$).
The amount of stars the outer disk is forming is comparable to the inner disk and it is a
significant ($\sim$10--15\%) fraction of the total star formation rate of the galaxy. This
clearly shows, that at least for M94, the outer disk is a very active region of the galaxy. This
activity is directly visible when looking at the UV and IR images shown in Fig. \ref{colmosaic},
where the unobscured young stars  as well as the hot dust are clearly seen in form of spiral
arms in the outer region of the galaxy.

Another important parameter to measure is the specific star formation rate (sSFR), i.e.
the amount of star formation per unit stellar mass. This value informs about the
efficiency of the star formation in the different regions of the galaxies. The sSFR can
be derived from the combination of Eqs. \ref{massdensity} and \ref{musfr} to give:

\begin{eqnarray}
\log(sSFR)(yr^{-1})=-0.4(FUV-m_{\lambda})-\log(M/L_{\lambda})
\nonumber\\
-0.4m_{abs,\sun,\lambda}-7.216
\label{sfr}
\end{eqnarray} 

\begin{figure}

\includegraphics[angle=0,scale=1]{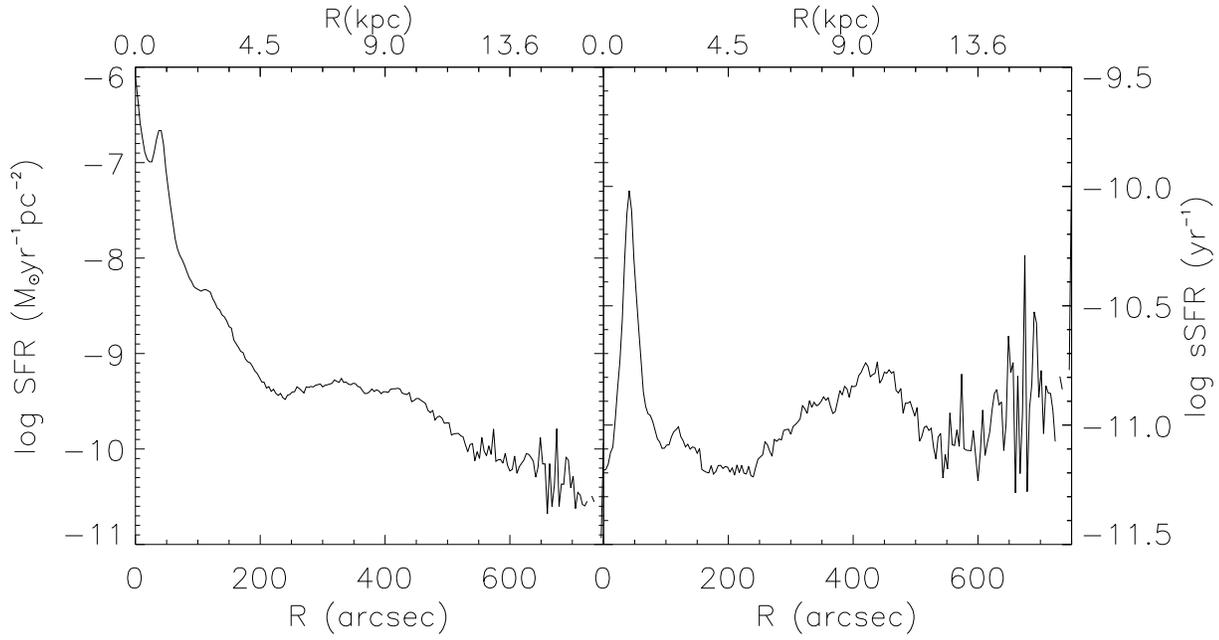}

\caption{Left panel: SFR profile of M94. Right panel. sSFR profile of M94.}
\label{sfrprof}

\end{figure}

Both the SFR and sSFR radial distribution are plotted in Fig. \ref{sfrprof}. Our
estimates agree very well by those produced by Boissier et al. (2007) for the star
formation rate density (their Fig.\ 9.39) of this galaxy: $\sim$10$^{-8}$$M_{\sun}pc^{-2}yr^{-1}$ at
50\arcsec (bulge), $\sim$5$\times$10$^{-9}$$M_{\sun}pc^{-2}yr^{-1}$ at 150\arcsec (inner
disk) and $\sim$4$\times$10$^{-10}$$M_{\sun}pc^{-2}yr^{-1}$ at 300\arcsec (outer disk).
Mu\~noz-Mateos et al. (2007) have noted that nearby spirals show a slightly positive
increase of the sSFR from inside-out. This result also applies in the particular case of
M94, where we find that  the outer disk has a larger sSFR ($\sim$0.012 Gyr$^{-1}$) than
the inner disk ($\sim$0.006 Gyr$^{-1}$). In other words, per unit stellar mass, the
outer disk is more active (efficient) forming stars.

\subsection{Spectral Energy Distributions}

\begin{figure}

\includegraphics[angle=0,scale=1]{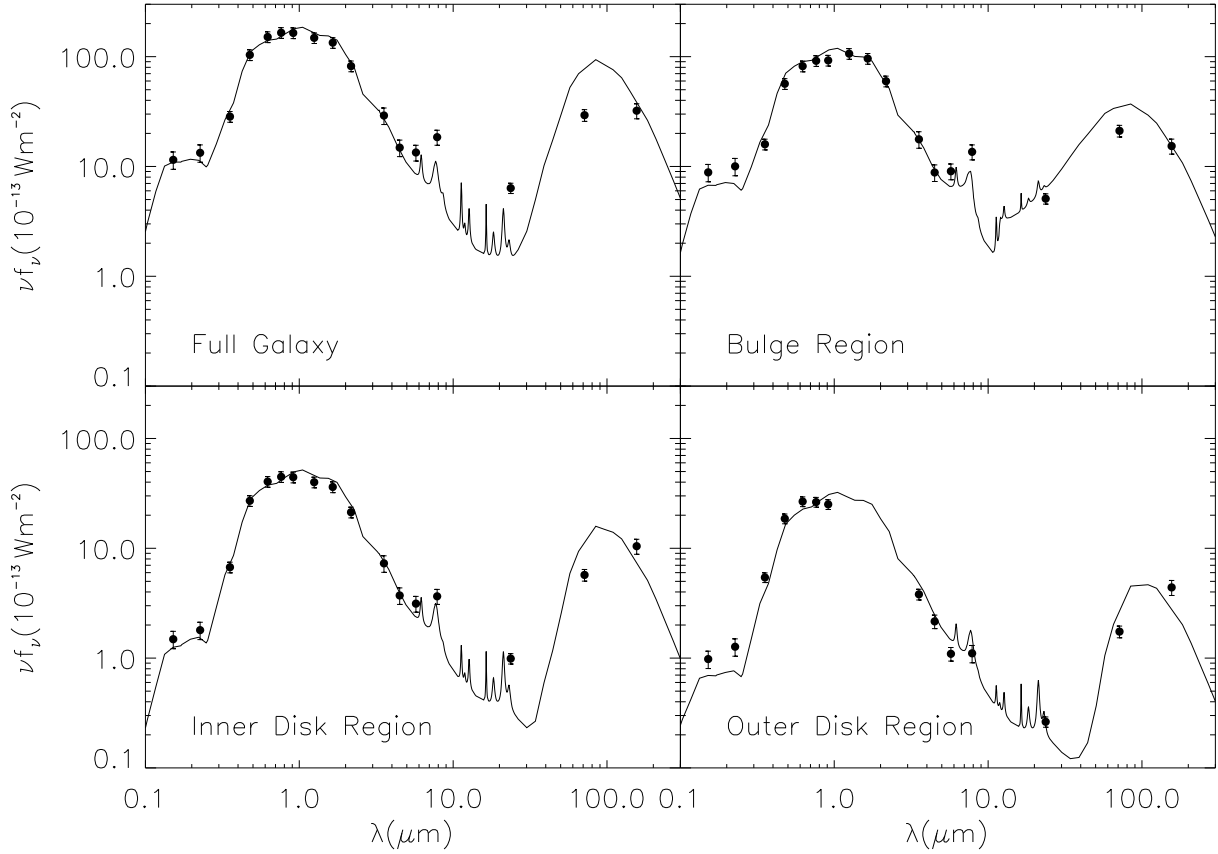}

\caption{M94 (NGC4736) galaxy 0.15--160$\mu$m SEDs of the total galaxy, its bulge,  inner disk  
and outer disk regions. The solid points are the observed data and the line is the best fit
model.}

\label{sedregions}

\end{figure}

Using the fluxes obtained in the 17 filter bands we have created the Spectral Energy
Distributions (SEDs) ranging from 0.15-160$\mu$m in the different regions of the galaxy
(see Table \ref{sedfluxes}). The SEDs across the galaxy are shown in  Fig.
\ref{sedregions}.

To extract some relevant physical parameters from the SEDs we have used synthetic templates
computed with the code GRASIL (Silva et al. 1998). This code couples the spectral (and chemical)
evolution of stellar populations with the radiative transfer through a dusty interstellar
medium. Consequently, the models consistently account for the emission from stars as well as
absorption and emission from dust in galaxies. GRASIL allows the variation of a large set of
free parameters but here we concentrate only in the small subset that mostly affect the shape of
the SED (Panuzzo et al. 2007). We describe briefly which parameters were left to change during
the fitting of the SEDs (see Silva et al. 1998 for a detailed description of the code):

\begin{enumerate}

\item Total infall mass, M$_{inf}$. This is the amount of primordial gas that has infalled to
form the galaxy. We have allowed the following range of variation depending on the region of the
galaxy which is fitted: bulge (0.8 to 5$\times$10$^{10}$M$_\sun$), inner disk (0.6 to
3$\times$10$^{10}$M$_\sun$), outer disk (0.6 to 3$\times$10$^{10}$M$_\sun$) and the full galaxy
(1 to 7$\times$10$^{10}$M$_\sun$). 

\item Exponential infall time scale, $\tau_b$. Gas is assumed to be continuously
infalling at a rate proportional to $\exp{(-t/\tau_b)}$. $\tau_b$ was allowed to change
from 0.1 to 10 Gyr.

\item Efficiency of Schmidt--type law, $\nu$. The SFR is assumed in GRASIL to follow a
Schmidt--type law of the form SFR(t)=$\nu$M$_g$(t)$^k$, where M$_g$ is the gas mass at any time,
k the exponent of the Schmidt law which we fixed to 1 , and $\nu$ is the free efficiency
parameter varying from 10$^{-3}$ to 4.

\item Fraction of molecular mass to total gas mass (M$_{mol}$/M$_{gas}$). We allow this
parameter to range from 0.1 to 0.99.

\item The escape timescale of newly born stars from their parent molecular clouds,
t$_{esc}$. We let this vary from 10$^{-3}$ to 10$^{-1}$ Gyr.

\item The mass of the molecular clouds, M$_{MC}$. The mass of the molecular clouds
together with the radii of the molecular clouds is related to the optical depth of the
molecular clouds, $\tau_{MC}\propto M_{MC}/r_{MC}^2$. We fixed r$_{MC}$=14 pc  and let 
M$_{MC}$ vary from 10$^{3}$ to 10$^{7}$ M$_{\sun}$.

\end{enumerate}

To allow a comparison with other works the adopted IMF is a Salpeter IMF in the mass
range from 0.1 to 100 M$_{\sun}$. Due to the low inclination of our galaxy we use the
output of the code with zero inclination. We assume that the galaxy has a
present age of 13 Gyr.

For the SED of each region we have computed the minimum $\chi^2$. Due to the large number of
free parameters (6 in our case) we have avoided a brute force approach that implies to compute a
large set of templates. For example, assuming that ten values per parameter are desired, 
10$^{N_{par}}$ templates should be created. In current powerful desktops every GRASIL template
takes $\sim$30 seconds to be computed. Consequently, the amount of time required in a brute
force approach is prohibitive. To overcome this difficulty, we have applied a Markov chain Monte
Carlo technique based on the Metropolis algorithm (Metropolis et al. 1953; Neal 1993). The
details of the code that we have used are in Asensio Ramos et al. (2007).

The best fit parameters together with the 1 $\sigma$ error are shown in Table
\ref{grasilfits}. Together with these values GRASIL also provides the present SFR and the
optical depth of molecular cloud at 1 $\mu$m $\tau_{MC}$ and the amount of mass in gas and  
stars.

We can conduct now a comparison between the stellar mass obtained using the (g-r) color and the
ones obtained using the GRASIL code. Comparing the stellar mass determined with GRASIL with the
ones derived in Section 6.2, following Bell et al. (2003), we find them systematically
($\sim$0.7) lower. However, it is worth stressing that within the error bars both estimations
agree. From the above comparison we think it is safe saying that the amount of stellar mass
contained within the different regions of the galaxy is as follows:
(2.8$<$M$_\star$$<$3.2)$\times$10$^{10}$$M_{\sun}$ (bulge),
(1.3$<$M$_\star$$<$1.7)$\times$10$^{10}$$M_{\sun}$ (inner disk),
(0.9$<$M$_\star$$<$1.5)$\times$10$^{10}$$M_{\sun}$ (outer disk), and
(4.2$<$M$_\star$$<$6.5)$\times$10$^{10}$$M_{\sun}$ (total). In relation to the SFR, we find
that GRASIL and the prescription presented in Mu\~noz-Mateos et al. (2007) agree very well
within the error bars. 

The GRASIL prediction in relation to the amount of present gas (i.e. not transformed into stars)
in the different regions of the galaxy can be also compared with the recent measurements of HI
(Walter et al. 2008). These authors find a total amount of gas of 4$\times$10$^{8}$$M_{\sun}$
for M94. This number is in excellent agreement with the GRASIL prediction
(4.5$\pm$1.0)$\times$10$^{8}$$M_{\sun}$. 

From the GRASIL results one can extract the following information: 1. The efficiency of the star
formation, as characterized by the $\nu$ parameter is very similar through the whole structure
of the galaxy, with no significant changes from the bulge to the outer disk. 2. The mass of the
typical molecular clouds where the star formation is taking place seems, however, to be much
smaller in the inner region of the galaxy than in the outer region. This could be related to the
much more violent conditions (influence of the active nucleus) close to the center of the galaxy
preventing the formation of massive molecular clouds. 3. The exponential infall time scale of
the gas seems to vary also across the galaxy. The gas appears to take more time to be deposited
in the bulge than what requires to be placed in the outer disk. This could be understood if the
gas requires, as expected, a large loss of angular momentum to feed the bulge than the inner
disk.

\section{Origin of the outer disk}

In the following we will address the origin of M94's bright outer star-forming disk. We will
explore the two obvious possibilities of it having an external versus internal origin.

\subsection{Merger origin} 

It has been proposed (Pe\~narrubia et al. 2006) that extended galactic disks can be form by a
tidally disrupted dwarf galaxy in prograde orbit that is coplanar with the host galaxy disk. As
a result of this merging, the stellar debris ends up resettling into an extended exponential
disk in dynamical equilibrium. Depending on the mass of the satellite, orbit eccentricity and
inclination, the stellar debris exhibit a rotation that could range from 30--50 km/s lower than
the host circular velocity up to a factor of $\sim$2 smaller. 

Favoring a merger origin for the  outer disk of M94 is the fact that its rotation curve is
declining from the inner region ($\sim$3 kpc) to the outer region ($\sim$9 kpc) by  $\sim$30\%
(Mulder 1995; de Blok et al. 2008). However, a common concern with the merging scenario
(particularly if the merger has a large mass ratio) is the possibility that the accretion
destroy the inner disk (Hopkins et al. 2009). In our case, the outer disk of M94 contains a
large fraction ($\sim$23\%) of the total mass of the galaxy. If all the outer disk were entirely
the result of a disrupted dwarf the implied merger ratio would be as high as 1:4. 


Another problem with the merger scenario, if we want to explain entirely the outer region of M94
as a tidally disrupted satellite, is the very different ellipticity the inner disk and the outer
region has. In the merging hypothesis the ellipticities of  both components can be interpreted
as a reflect of the different inclination they have. In fact, transforming the observed
ellipticities to inclinations, we find that the inner region will have an inclination of
$\sim$35 degrees whether the outer disk will have only $\sim$8 degrees.  Merging simulations of
massive satellites with disk galaxies whereas the disk survived the encounter only  found a
maximum tilt of $\sim$7 degrees for the outer regions (Vel\'azquez \& White 1999). This is much
smaller than what is observed for this galaxy.

Even in the case that the origin of the outer region of M94 could not be explained by a major
merger it could be well the case that an interaction of M94 with another galaxy could be the
responsible of triggering both the star formation we see in this part of the galaxy as well as
the many morphological structures this galaxy has: inner ring, inner disk/oval distortion, gap,
outer spiral arms, etc. The interaction possibility is however disfavored by the observations.
Although M94 is the main member of the Canes Venatici I Group, it seems unlikely that it has
suffered any interaction with another member of that group recently. In fact, such group has
been found to be only weakly gravitationally bound. Moreover, most of the galaxies in this group
appear to be moving with the expansion of the universe, which will prevent nearby (repeated)
encounters affecting M94 (Karachentsev et al. 2003).

\subsection{Secular evolution}

We also explore whether the plethora of structures we see in M94 as well as the enhanced star
formation in its periphery could be created assuming a perturbation of internal (secular)
origin. It is well known that bars can be produced via dynamical instabilities in galactic
disks. A particular form of bars is what is known as oval distortion which are weak bars also
called "fat bars" that often resemble normal slightly inclined disk having even spiral arms (see
e.g. Jogee et al. 2002). As mentioned in Sec. 2, the inner disk of M94 has been sometimes
considered as an oval distortion more than a truly inclined disk. This seems to be a  reasonable
hypothesis since the inclination of the  galaxy, as derived in the UV and IR imaging for the
outer disk, is lower than what it would be derived from the shape of the inner disk.  It is
worth stressing that this conclusion suggesting that the inner region of M94 is an oval
distortion more than an truly inclined disk can not firmly established on the basis of the
optical images alone (which were the main source of information used in previous works), but,
both the UV and IR images presented in this paper clearly show that there is a continuation in
the spiral arms from the inner (oval) region to the outer disk. If both structures were
decoupled, as suggested by the apparent different inclination and explained in the merger
scenario, this prolongation of the spiral arms would be extremely unlikely. Under the hypothesis
that the inner region is an oval distortion the problem with the different inclinations among
the inner and the outer disk of M94 is solved since the oval distortion is not inclined but has
a different ellipticity than the outer disk because is intrinsically oval and not circular.

If the inner region is an oval distortion, this structure could be a serious candidate to
explain the spiral arm structure we see in the outer disk (Binney \& Tremaine 1987, and
references therein), and consequently to be the ultimate responsible of the enhanced star
formation we measure in this part of the galaxy. Oval distortions are known to create spiral
arms that can be sustained for long time in isolated galaxies (e.g., Lindblad 1960; Toomre
1969; Sanders \& Huntley 1976; Athanassoula 1980). In fact, oval distortions have been
considered as important as bars when considering the secular evolution of a disk galaxy
(Kormendy \& Kennicutt 2004). 

\begin{figure}
\begin{center}
\includegraphics[scale=0.8,angle=-90]{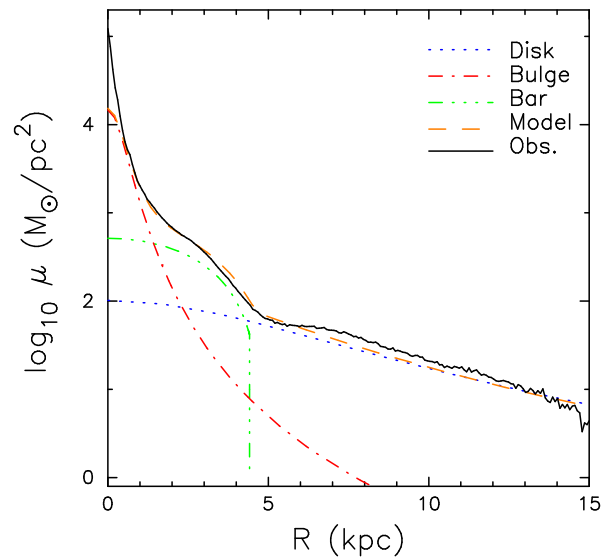}

\caption{Fit to the surface density profile (shown in Fig. 4) using different analytical
components. The oval distortion (inner disk) is modeled with a Ferrers bar with n=1. The bulge,
outer disk and halo are fit using Miyamoto-Nagai models. }

\label{fig:suden}
\end{center}
\end{figure}

We have created a toy model to explore whether the observed oval distortion can produce
spiral arms similar to what we see in the outer disk. Our simple model is  composed of a
bulge, an oval distortion, an outer disk and a dark matter halo. To characterize the
different components of the galaxy we use different analytical formulas that we fit to
the stellar mass density profile in order to have realistic parameters (see Fig.
\ref{fig:suden}). Considering the oval distortion as a very mild bar, we use the
analytical formula by Ferrers (1877) to describe it:

\begin{equation}
\begin{centering}
\rho=\left\{\begin{array}{ll}
\frac{15}{8}\frac{M_{od}}{\pi abc}(1-m^2)^1, & m\le1\\
0, & m\ge1
\end{array}\right.
\end{centering}
\label{eqn:Ferrer}
\end{equation}

where $m^2=\frac{x^2}{a^2}+\frac{y^2}{b^2}+\frac{z^2}{c^2}$, and $a$ is the major axis, $b$ the
minor, and $c$ the vertical axis, taken here as $500$~pc, with $a>b>c$. This bar model is very
popular in the literature (see e.g. Sanders \& Tubbs 1980; Papayannopoulos \& Petrou 1983;
Schwarz 1985; Athanassoula 1992; Patsis 2005, etc). From the fit to the stellar surface mass density we
obtain: a=$4.9$~kpc, b=$4.1$~kpc and a total mass for the oval distortion  equal to
$M_{od}=1.3\times10^{10}M_{\sun}$. 

The remaining components of the galaxy (the bulge, the outer disk and the dark matter
halo) have been modeled using the Miyamoto \& Nagai (1975) analytical models. The
Miyamoto-Nagai potential is described by the equation:

\begin{equation}
\begin{centering}
\Phi_{M}(R,z)=-\frac{GM}{\sqrt{R^2+(a+\sqrt{z^2+b^2})^2}}
\end{centering}
\label{eqn:MiyNag}
\end{equation}

\begin{figure}
\begin{center}
\includegraphics[scale=0.4,angle=0]{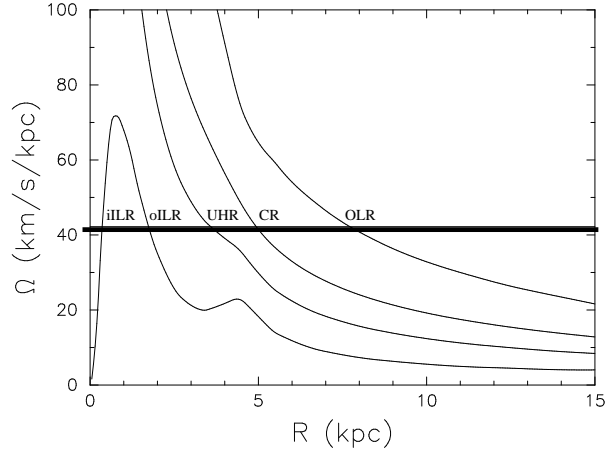}

\caption{Angular frequency of the proposed model. The solid horizontal line is the pattern
speed chosen in this paper.}

\label{fig:axikap}
\end{center}
\end{figure}

The parameters retrieved from the fit to the stellar mass surface density are: bulge
(M=2.0$\times$10$^{10}$M$_\sun$, $a=0.05$ kpc and $b=0.6$ kpc); outer disk
(M=2.8$\times$10$^{10}$M$_\sun$, $a=6.5$ kpc and $b=0.5$ kpc) and halo
(M=20.0$\times$10$^{10}$M$_\sun$, $a=0.0$ kpc and $b=30$ kpc). Note that the outer disk in this
model has a factor of $\sim$2 more stellar mass than what we have measured before using the
outer region (200--430\arcsec) of the stellar mass profile. This is explained because in the
model the stellar mass density profile is extrapolated in- and outwards, to zero and infinity
respectively. Once the galactic potential is set, we can start running a simulation and see how
a gaseous disk responds to the potential. Before doing this we can use the linear approximation
to calculate the angular frequency, and therefore estimate the radii of the main resonances
(Fig.~\ref{fig:axikap}) associated to the gravitational potential we have created. This is
important for this galaxy since several resonant structures like the inner ring are observed.

One of the key issues when considering bars, or oval distortions, and their corresponding
resonances is the pattern speed, i.e., how fast the bar rotates as a solid body. The pattern
speed determines, many different structures, such as the radii of the different resonant rings,
or their actual existence. In the case of our study, the variation of the pattern speed, also
determines the appearance of the spiral arms and their shape. A precise study of the spiral arms
and their origin (Romero-G\'omez et al. 2007), establishes that when increasing the pattern
speed in a given barred model, the spiral arms become more tightly wounded.

Several pattern speeds have been proposed to explain different features in M94. We started with
$\Omega_{p}$=$56$~km~s$^{-1}$~kpc$^{-1}$ proposed by Mulder \& Combes (1997) and decreased this
value down to $\Omega_{p}$=$30$~km~s$^{-1}$~kpc$^{-1}$. We found that a model that is able to
reproduce the shape of the outer spiral arms as well as the inner ring of the galaxy has
$\Omega_{p}$=$38$~km~s$^{-1}$~kpc$^{-1}$. This number agrees nicely with previous studies, such
as Waller et al. (2001), where they studied the central part of the galaxy and obtained
$\Omega_{p}$=$35$~km~s$^{-1}$~kpc$^{-1}$. With the pattern speed that we are using,
R$_{CR}$~=~5.2~kpc, and consequently R$_{CR}$/R$_{bar}$=1.04. In that sense, the oval distortion
that we have created  would be considered as a fast bar (Aguerri et al. 2003). In addition, our
pattern speed  produces  an inner ring at a radius of $\sim$1 kpc, at the same distance than the
one observed.

\begin{figure}
\begin{center}
\includegraphics[scale=0.7,angle=-90]{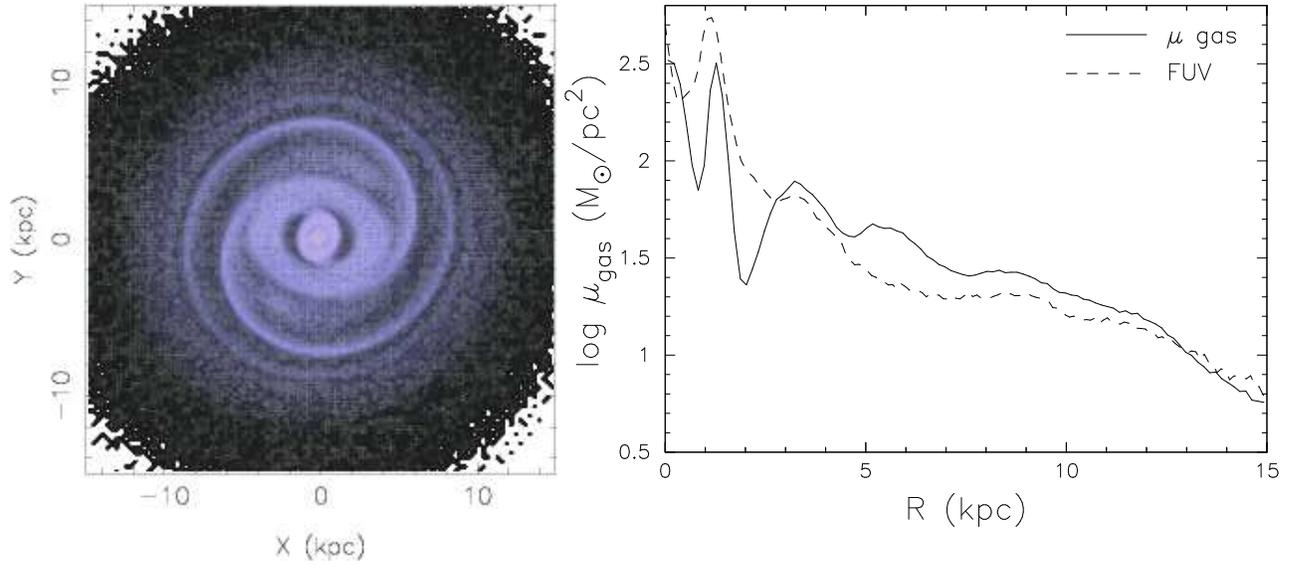}
\includegraphics[scale=0.5,angle=-90]{fig9b.ps}

\caption{{\it Left panel}: Snapshot of the SPH simulation showing the spiral arms as a
response to the inner oval distortion. On the {\it right panel}, the solid line shows the gas
surface density profile of our model. The different peaks represent the response of the gas
to the dynamical effect of the oval distortion. The dashed line shows the FUV  surface
brightness (i.e. a good proxy to the recent star formation) scaled to fit in the plot.}

\label{fig:simu}
\end{center}
\end{figure}

Once the gravitational potential and the pattern speed are fixed we run our simulation to
explore the response on a pure gas disk. We use the code FTM 4.4 (updated version) from
Heller \& Shlosman(1994), which allow us to study the dynamical response of the gas based on
SPH. We use 150000 collisional particles to simulate the gas, isothermal and non
self-gravitating. Initially the gaseous particles are distributed following a Miyamoto-Nagai
density distribution. In Fig.~\ref{fig:simu}, {\it{left panel}}, we show a snapshot of our
simulation showing clearly the spiral arms, as a result of the oval distortion. In addition,
the inner ring that is observed in the real galaxy can be seen. To make a more quantitative
comparison of our toy model with the observations, in Fig. \ref{fig:simu}, {\it{right
panel}}, we compare the density profile of the gas in our simulation with the FUV surface
brightness profile. The peaks on the FUV profile represents the places where the star
formation is enhanced. We can see that there is a reasonable agreement between the peaks in
the FUV and the position of the peaks where the density of the gas in the model is higher.

According to our simulation,  if the oval distortion survives for long time (Athanassoula \&
Misiriotis 2002, Martinez-Valpuesta, Shlosman \& Heller 2006) the spiral arm and the ring will
also survive. However, a slow secular evolution is expected in the radius of the oval structure
and the pattern speed. In addition, we should take into account the slowly consumption of the
gas by the star formation (which could be somehow counterbalance by the new gas infall), that
will also decrease the amount of gas in the spiral arms and oval distortion.

In a following paper (Mart\'inez-Valpuesta et al., in preparation) we plan to increase the
resolution of our hydrodynamical simulation to explore in more detail the inner region of the
galaxy. In particular, we will explore the velocity shear field of the galaxy to probe whether the
dynamics linked to the oval distortion can help to explain the GRASIL findings in relation to the
efficiency, molecular cloud mass and infall time scale found from the stellar population analysis.
Our aim is to find a consistent model able to reproduce at the same time the new HI kinematic data
(Blok et al. 2008) in the outer region.

\section{Conclusions}

M94 is the closest example of an early--type spiral galaxy with low inclination. It has been
traditionally considered (due to its optical appearance) as a galaxy with a bright outer ring
surrounding its disk. The multiwavelength analysis that we have conducted in this paper show
that this view is incorrect. Observed in the UV and in the IR the galaxy reveals that its outer
region has a disk with a spiral arms structure nature. This outer disk is very active forming
stars and contains $\sim$23\% of the total mass of the galaxy. We have explored what is
enhancing this star formation activity in the galaxy periphery and find that a likely candidate
is the dynamical effect of the inner oval distortion on the rest of the galaxy. Although an
external origin of the extended disk (e.g. due to the accretion of a satellite galaxy) can not
be ruled out with the existing data, the oval distortion hypothesis explains more naturally the
existence of an inner ring and the enhanced star formation activity in the outer disk of M94. In
addition, it is likely that the flux of gas that the oval distortion produces towards the center
of the galaxy would be related to the active nature of the bulge of this galaxy.


\acknowledgments

We are happy to thank Jose Antonio Acosta Pulido, Juan Carlos Mu\~noz-Mateos, Armando Gil de Paz and
Olga Vega for their interesting comments on different aspects of this work. Thanks are also given to
Andres Asensio Ramos and Jose Alberto Rubi\~no Mart\'in for their help on using their Markov chain
algorithm, and to Yoshiaki Sofue. We also would like to thank to the THINGS team (especially Erwin
de Blok) for providing us with kinematical data of M94. We also acknowledge the detailed review of
the manuscript done by the referee. IT and DM-G acknowledge support from the Ram\'on y Cajal Program
financed by the Spanish Government.



\clearpage

\begin{deluxetable}{ccccc}
\tabletypesize{\scriptsize}
\tablecaption{Data quality and galactic extinction correction}
\tablewidth{0pt}
\tablehead{
\colhead{Band} & \colhead{$\mu_{crit}$(mag/arcsec$^2$)} & \colhead{R$_{crit}$(\arcsec)}
 & \colhead{PSF FHWM (\arcsec)} & \colhead{A$_\lambda$ (mag)}}
\startdata
FUV & 31.5 & 550  & 4.5 &   0.134   \\
NUV & 31.3 & 650  & 4.7 &  0.119 \\
u & 28.2 & 660 & 1.5  & 0.093 \\
g & 28.2 & 710 &   1.2  &  0.070  \\
r & 27.5 & 730  &  1.0  & 0.050 \\
i & 27.2 & 690 & 1.1  & 0.037 \\
z & 26.7 & 670  &  1.0  & 0.030   \\
J & 24.1 & 205  &  3.0  & 0.017 \\
H & 23.8 & 210 &  2.9 &  0.010 \\
K & 24.3 & 210   &   3.0  & 0.007 \\
3.6$\mu$m & 26.6 & 660  &  1.7  & 0.004 \\
4.5$\mu$m & 25.8 & 630 & 1.9  & 0.001 \\
5.8$\mu$m & 25.0 & 360 & 2.0   & \nodata \\
8.0$\mu$m & 26.0 & 415 & 2.2  & \nodata \\
24$\mu$m & 25.5 & 360 & 5.5$^a$  & \nodata \\
70$\mu$m & 23.2 & 410  & 15.8$^a$  & \nodata \\
160$\mu$m & 21.5 & 450 & 34.6$^a$ & \nodata

\enddata

\tablecomments{Last column, A$_\lambda$, is the magnitude correction applied to correct for
Galactic extinction (Schlegel et al. 1998) using E(B-V)=0.018 mag.}

\tablenotetext{a}{Values obtained from a Gaussian fit to theoretical PSF models constructed
by the Tiny Tim Point Spread Function modeling program (developed for the Spitzer Science
Center by John Krist; STScI) to a blackbody source of 50K temperature
(http://ssc.spitzer.caltech.edu/mips/psf.html).}

\label{surcrit}

\end{deluxetable}

\begin{deluxetable}{lccccccccccccccccc}
\tabletypesize{\scriptsize}
\tablecaption{Ultraviolet, Optical, Near-Infrared, Infrared flux densities}
\tablewidth{0pt}
\tablehead{
\colhead{Rad. Ran.} & \colhead{FUV} & \colhead{NUV}
 & \colhead{u} & \colhead{g} & \colhead{r} & \colhead{i} & \colhead{z}
 & \colhead{J} & \colhead{H} & \colhead{K$_S$} & \colhead{3.6} 
 & \colhead{4.5} & \colhead{5.8} & \colhead{8} & 
 \colhead{24} & \colhead{70} & \colhead{160}    \\
\colhead{(\arcsec)} & \colhead{(Jy)} & \colhead{(Jy)}
 & \colhead{(Jy)} & \colhead{(Jy)} & \colhead{(Jy)} & \colhead{(Jy)} & \colhead{(Jy)}
 & \colhead{(Jy)} & \colhead{(Jy)} & \colhead{(Jy)} & \colhead{(Jy)} 
 & \colhead{(Jy)} & \colhead{(Jy)} & \colhead{(Jy)} & 
 \colhead{(Jy)} & \colhead{(Jy)} & \colhead{(Jy)} }

\startdata
0 -- 75 & 0.045 &   0.076  &   0.19 &    0.90  &    1.70  &    2.33
   &   2.82   &   4.44   &   5.28  &    4.32   &   2.09   &   1.32
   &   1.73   &   3.57   &   4.02  &    $>$50.29   &   $>$79.72 \\
75 -- 200 &    0.008  &  0.014 &   0.08  &   0.43  &   0.84  &    1.14
   &   1.35 &     1.67   &   1.99  &    1.54   &  0.86  &   0.56
   &  0.60 &    0.96  &   0.78  &    $<$13.62   &   $<$54.33 \\
200 -- 430 &   0.005 &  0.010  &  0.06   &  0.30  &   0.55 &    0.67
   &  0.77  &   \nodata  &   \nodata  &   \nodata   &  0.45  &   0.32
   &  0.21  &   0.29  &   0.21  &    $<$4.15   &  $<$22.88 \\
0 -- 430 &  0.058   &  0.100   &  0.34   &   1.65  &    3.14  &    4.20
   &   5.01  &    $>$6.18 &    $>$7.37 &    $>$5.94 &     3.44   &   2.22
   &   2.56 &     4.85  &    5.02  &    69.85   &   167.34 \\

\enddata



\label{sedfluxes}

\end{deluxetable}

\begin{deluxetable}{lcccccccccc}
\tabletypesize{\scriptsize}
\tablecaption{Best-fit parameters and derived quantities of the SED GRASIL models}
\tablewidth{0pt}
\tablehead{
\colhead{Rad. Ran.} & \colhead{$\tau_b$} & \colhead{M$_{inf}$}
 & \colhead{$\nu_{sch}$} & \colhead{M$_{mol}$/M$_{gas}$} & \colhead{t$_{esc}$} 
 & \colhead{M$_{MC}$} & \colhead{SFR} & \colhead{$\tau_{MC}$} & \colhead{M$_{gas}$} & \colhead{M$_{\star}$} \\
\colhead{(\arcsec)} & \colhead{(Gyr)} & \colhead{(10$^{10}$M$_\sun$)}
 & \colhead{(Gyr$^{-1}$)} & \colhead{} & \colhead{(Gyr)} 
 & \colhead{(10$^{6}$M$_\sun$)} & \colhead{(M$_\sun$ yr$^{-1}$)} & \colhead{} & \colhead{(10$^{8}$M$_\sun$)} & \colhead{(10$^{10}$M$_\sun$)} }

\startdata
0 -- 75    & 3.7$\pm$0.9 & 3.1$\pm$0.5 & 2.4$\pm$0.8 & 0.49$\pm$0.25 &  0.030$\pm$0.017 & 1.3$\pm$0.7 & 0.64$\pm$0.40 & 60$\pm$14   & 2.63$\pm$0.62 & 2.75$\pm$0.45 \\
75 -- 200  & 2.8$\pm$0.9 & 1.4$\pm$0.3 & 2.1$\pm$1.0 & 0.52$\pm$0.24 &  0.051$\pm$0.023 & 6.9$\pm$2.9 & 0.19$\pm$0.18 & 346$\pm$98  & 0.84$\pm$0.31 & 1.27$\pm$0.30 \\
200 -- 430 & 2.2$\pm$0.6 & 1.0$\pm$0.1 & 2.3$\pm$0.8 & 0.71$\pm$0.14 &  0.032$\pm$0.009 & 8.9$\pm$1.3 & 0.07$\pm$0.10 & 486$\pm$12  & 0.28$\pm$0.26 & 0.86$\pm$0.62 \\
0 -- 430   & 3.9$\pm$1.1 & 4.8$\pm$0.8 & 2.3$\pm$0.8 & 0.49$\pm$0.25 &  0.031$\pm$0.019 & 4.4$\pm$3.4 & 1.07$\pm$0.62 & 205$\pm$121 & 4.50$\pm$1.00 & 4.22$\pm$0.63 \\

\enddata



\label{grasilfits}

\end{deluxetable}


\begin{thebibliography}{}

\bibitem[]{} Aguerri, J.~A.~L., Debattista, V.~P., \& Corsini E.~M. 2003, \mnras, 238, 465
\bibitem[]{} Asensio Ramos, A., Mart\'inez Gonz\'alez, M. J. \& Rubi\~no-Mart\'in, J.
A., 2007, A\&A, 476, 959 
\bibitem[]{} Athanassoula, E., 1980, A\&A, 88, 184
\bibitem[]{} Athanassoula, E., 1992, \mnras, 259, 328
\bibitem[]{} Athanassoula, E. \& Misiriotis, A., 2002, \mnras, 330, 35
\bibitem[]{} Bakos, J., Trujillo, I., \& Pohlen, M., 2008, \apj, 683, L103
\bibitem[]{} Beckman, J. E., Varela, A. M., Mu\~noz-Tu\~n\'on, C., Vilchez, J. M.,
\& Cepa, J. 1991, A\&A, 245, 436
\bibitem[]{} Bell, E. F., McIntosh, D. H., Katz, N., \& Weinberg M. D., 2003, \apjs, 149,
289 
\bibitem[]{} Bianchi, L., et al. 2003a, in The Local Group as an Astrophysical Laboratory,
 ed. M. Livio \& T. Brown (Baltimore: STScI), 10
\bibitem[]{} Bianchi, L., et al. 2003b, BAAS, 203, 91.12
\bibitem[]{} Binney, J., \& Tremaine, S., 1987, Galactic dynamics (Princeton: Princeton Univ. Press)
\bibitem[]{} Bland-Hawthorn, J., Vlaji\'c, M., Freeman, K. C., \& Draine, B. T.,
2005, \apj, 629, 239
\bibitem[]{} Boissier, S. et al., 2007, ApJS, 173, 524
\bibitem[]{} Bosma, A., van der Hulst, J. M., \& Sullivan, W. T., III., 1977, A\&A, 57, 373
\bibitem[]{} Bournaud, F., Elmegreen, B. G., \& Elmegreen, D. M., 2007, \apj, 670, 237
\bibitem[]{} Braun, R., 1995, A\&AS, 114, 409
\bibitem[]{} Buat, V., et al., 2005, ApJ, 619, L51
\bibitem[]{} Buta, R., 1988, \apjs, 66, 233
\bibitem[]{} Buta, R. J., Corwin H. G., Jr, Odewahn S. C., 2007, in The de Vaucouleurs
Atlas of Galaxies. Cambridge Univ. Press, Cambridge
\bibitem[]{} Cho, J., \& Park, C., 2009, \apj, 693, 1045
\bibitem[]{} Dale, D.A., et al.,2007, \apj, 655, 863
\bibitem[]{} Davies, E., 1990, in Machine Vision: Theory, Algorithms and Practicalities, Academic
Press, pp. 42 
\bibitem[]{} Debattista, V. P., Mayer, L., Carollo, C. M., Moore, B., Wadsley, J.,
\& Quinn, T., 2006, \apj, 645, 209
\bibitem[]{} de Blok, W. J. G., Walter, F., Brinks, E., Trachternach, C., Oh, S.-H.,
 Kennicutt, R. C., 2008, \aj, 136, 2648
\bibitem[]{} de Jong, R. \& Bell, E. F., 2006, astro-ph/0604391
\bibitem[]{} Donas, J., Milliard, B., Laget, M.  \& Deharveng, J. M., 1981, A\&A, 97, L7 
\bibitem[]{} de Vaucouleurs, G., de Vaucoleurs, A., Corwin, H. G., Jr., Buta, R.
J., Paturel, G., \& Fouqu\'e, P., 1991, Third Reference Catalogue of Bright Galaxies
(New York: Springer)
\bibitem[]{} Elmegreen, B. G. \& Hunter, D. A., 2006, \apj, 363, 712
\bibitem[]{} Elmegreen, B. G. \& Parravano, A., 1994, \apj, 435, L121
\bibitem[]{} Erwin, P., 2004, A\&A, 415, 941
\bibitem[]{} Erwin, P., Beckman, J. E., \& Pohlen, M., 2005, \apj, 626, L81
\bibitem[]{} Erwin, P., Pohlen, M., Gutierrez, L., Beckman, J. E., 2007, arXiv:0712.1473
\bibitem[]{} Erwin, P., Pohlen, M., \& Beckman, J. E., 2008, \aj, 135, 20
\bibitem[]{} Ferguson, A. M. N., \& Clarke, C. J., 2001, \mnras, 325, 781
\bibitem[]{} Ferrers, N.~M., 1877, Q.~J.~Appl.~Math., 14, 1
\bibitem[]{} Foyle, K., Courteau, S., Thacker, R. J., 2008, \mnras, 386, 1821
\bibitem[]{} Freeman, K. C., 1970, \apj, 160, 811
\bibitem[]{} Karachentsev, I. D., et al., 2003, A\&A, 398, 467
\bibitem[]{} Karachentsev, I. D., 2005, \aj, 129, 178
\bibitem[]{} Kazantzidis, S., Bullock, J. S., Zentner A. R., Kravtsov, A. V., \&
Moustakas, L. A., 2008, \apj, 688, 254
\bibitem[]{} Kennicutt, R. C., 1989, \apj, 344, 685
\bibitem[]{} Kennicutt, R. C., et al., 2003, PASP, 115, 928
\bibitem[]{} Kennicutt, R. C., 1998, ARA\&A, 36, 189
\bibitem[]{} Kormendy, J., \& Kennicutt, R.~C., 2004, ARA\&A, 42, 603
\bibitem[]{} Gil de Paz, A., et al.,  2004, BAAS, 205, 42.01
\bibitem[]{} Gil de Paz, A., et al.,2005, \apj, 627, L29
\bibitem[]{} Gil de Paz, A., et al., 2007, \apjs, 173, 185
\bibitem[]{} Haralick, R. \& Shapiro L., in Computer and Robot Vision, 
Addison-Wesley Publishing Company, 1992, Vol. 1, Chap. 7.
\bibitem[Heller \& Shlosman(1994)]{1994ApJ...424...84H} Heller, C.~H.,\& Shlosman, I.\ 1994, \apj, 424, 84
\bibitem[]{} Hopkins, P.F., Cox, T.J., Younger, J. D., \& Hernquist, L., 2009, \apj, 691, 1168
\bibitem[]{} Hunter, D. A., \& Elmegreen, B. G., 2005, \apjs, 162, 49
\bibitem[]{} Jarrett, T.H., Chester, T., Cutri, R., Schneider, S.,  
Huchra J. P., 2003, \aj, 125, 525
\bibitem[]{} Jogee, S., Knapen, J. H., Laine, S., Shlosman, I., Scoville, N. Z., Englmaier, P.,
2002, \apj, 570, L55
\bibitem[]{} Larkin, J. E., Armus, L., Knop, R. A., Soifer, B. T., Matthews, K., 1998, \apjs, 114,
59
\bibitem[]{} Lindblad, P.~0., 1960, Stockkholms Observatorium Ann., 21,4
\bibitem[]{} Maoz, D., Filippenko, A. V., Ho, L. C., Rix H.-W., Bahcall, J. N., Schneider, D. P.,
Macchetto, F. D., 1995, \apj, 440, 91 
\bibitem[]{} Martin, D. C., et al. 2005, ApJ, 619, L1
\bibitem{}{} Mart\'inez-Delgado, D., Pe\~narrubia, J., Gabany, R. J., Trujillo, I., Majewski,
S. R., \& Pohlen, M., 2008, \apj, 689, 184
\bibitem[]{} Martinez-Valpuesta, I., Shlosman, I., \& Heller, C.~H., 2006, \apj, 637, 214
\bibitem[]{} Metropolis, N., Rosenbluth, A. W., Rosenbluth, M. N., Teller, A. H., \&
Teller, E., 1953, J. Chem. Phys., 21, 1087
\bibitem[]{} Miyamoto, M., \& Nagai, R., 1975, \pasj, 27, 533
\bibitem[]{} M\"ollenhoff, C., Matthias, M., \& Gerhard, O. E. 1995, A\&A, 301, 359
\bibitem[]{} Mulder, P.S., 1995, A\&A, 303, 57
\bibitem[]{} Mulder, P.~S. \& Combes, F., 1996, A\&A, 313, 723
\bibitem[]{} Mu\~noz-Mateos, J.C., Gil de Paz, A., Boissier, S., Zamorano, J., Jarrett, T.,
Gallego, J., \& Madore, B. F., 2007, 658, 1006
\bibitem[]{} Mu\~noz-Tu\~n\'on, C., Prieto, M., Beckman, J. E., \& Cepa, J. 1989,
Ap\&SS, 156, 301
\bibitem[]{} Neal, R. M., 1993, Probabilistic Inference Using Markov Chain Monte Carlo
Methods (Dept. of Statistics, University of Toronto: Technical Report No. 0506)
\bibitem[]{} Panuzzo, P., et al., 2007, \apj, 656, 206
\bibitem[]{} Papayannopoulos, T., Petrou, M., 1983, A\&A, 119, 21
\bibitem[]{} Patsis, P. A., 2005, \mnras, 358, 305
\bibitem[]{} Pe\~narrubia, J., McConnachie, A., \& Babul, A., 2006, \apj, 650, L33
\bibitem[]{} Pohlen, M., Dettmar, R.-J.,  L\"utticke, R., \& Aronica, G., 2002,
A\&A, 392, 807
\bibitem[]{} Pohlen, M., \& Trujillo, I., 2006, A\&A, 454, 759
\bibitem[]{} Roberts, T.P., Warwick R. S. \& Ohashi, T., 1999, \mnras, 304, 52 
\bibitem[]{} Roberts, T.P., Schurch N. J. \&  Warwick R. S., 2001, \mnras, 324, 737
\bibitem[]{} Romero-G\'omez, M., Athanassoula, E., Masdemont, J.~J., \& Garc\'ia-G\'omez, C., 2007, A\&A, 472, 63
\bibitem[]{} Ro\v skar, R., Debattista, V. P., Stinson, G. S., Quinn, T. R.,
Kaufmann, T., Wadsley J., 2008, \apj, 675, L65
\bibitem[]{} Salpeter, E. E., 1955, \apj, 121, 161
\bibitem[]{} Sanders, R.~H. \& Huntley, J.~M., 1976, \apj, 209, 53
\bibitem[]{} Sanders, R. H., \& Tubbs, A. D., 1980, \apj, 235, 803
\bibitem[]{} Shaw, M. A.,  Combes, F., Axon, D. J., \& Wrigth, G. S., 1993, A\&A, 273, 31 
\bibitem[]{} Schaye, J., 2004, \apj, 609, 667
\bibitem[]{} Schlegel, D. J., Finkbeiner, D. P., Davis, M., 1998, \apj, 500, 525
\bibitem[]{} Schwarz, M. P., 1985, \mnras, 212, 677
\bibitem[]{} Silva, L., Granato, G. L., Bressan, A., \& Danese, L., 1998, \apj, 509, 103
\bibitem[]{} Smith, B. J., Lester, D. F., Harvey, P. M., Pogge, R. W., 1991, \apj, 373, 66
\bibitem[]{} Smith, B. J., Harvey, P. M., Colom\'e, C., Zhang, C. Y., DiFrancesco, J., Pogge, R. W.,
1994, \apj, 425, 91
\bibitem[]{} Slyz, A. D., Devriendt, J. E. G., Silk, J., \& Burkert, A., 2002,
\mnras, 333, 894
\bibitem[]{} Sofue, Y., Tutui, Y., Honma, M., Tomita, A., Takamiya, T., Koda, J., \& Takeda,
Y., \apj, 1999, 523, 136
\bibitem[]{} Taniguchi, Y., Ohyama, Y., Yameda, T., Mouti, H., Yoshida, M., 1996, \apj, 467, 215
\bibitem[]{} Thilker D. A., et al., 2005, \apj, 619, L79
\bibitem[]{} Toomre, A., 1969, \apj,  158, 899
\bibitem[]{} Turner, J. L. \& Ho, P.T.P,  1994, \apj, 421, 122 
\bibitem[]{} van den Bosch, F. C., 2001, \mnras, 327, 1334
\bibitem[]{} van der Kruit, P. C., 1976, A\&A, 52, 85
\bibitem[]{} van der Kruit, P. C., 1979, A\&AS, 38, 15
\bibitem[]{} van der Kruit, P. C., 1987, A\&A, 173, 59
\bibitem[]{} van der Kruit, P. C., \& Searle, L., 1981a, A\&A, 95, 105
\bibitem[]{} van der Kruit, P. C., \& Searle, L., 1981b, A\&A, 95, 116
\bibitem[]{} Waller, W.~H. et al., 2001, AJ, 121, 1395
\bibitem[]{} Walter, F., Brinks, E., de Blok, W. J. G., Bigiel, F. Kennicutt, R. C., 
Thornley, M. D., Leroy, A., 2008, \aj, 136, 2563
\bibitem[]{} York, D. G., et al., 2000, \aj, 120, 1579
\bibitem[]{} Yoshii, Y., \& Sommer-Larsen, J., 1989, \mnras, 236, 779
\bibitem[]{} Younger, J. D., Cox, T. J., Seth A. C., \& Hernquist, L., 2007, \apj, 670, 269
\bibitem[]{} Younger, J. D., Besla, G., Cox, T. J., Hernquist, L., Robertson, B., \& Willman,
B., 2008, \apj, 676, L21

\end{thebibliography}
\end{document}